\documentclass[fleqn,10pt]{article}

\usepackage{arxiv}
\usepackage{{jabbrv}}
\usepackage{cite}
\usepackage{filecontents}
\usepackage{titletoc}

\usepackage[utf8]{inputenc}
\usepackage[T1]{fontenc}

\usepackage{gensymb}

\usepackage{hyperref}

\usepackage{color}

\usepackage{amsmath,amsfonts,amssymb}
\usepackage{graphicx}
\usepackage{booktabs}

\usepackage[left=2cm,right=2cm,top=2.25cm,bottom=2.25cm,headheight=12pt,letterpaper]{geometry}

\title{Chaining of hard disks in nematic needles: 
  particle-based simulation of colloidal interactions in liquid crystals} 

\author{
  David M\"uller \\
  Physics Department \\
  TU Dortmund University \\
  Dortmund 44227 Germany \\
  \texttt{david2.mueller@udo.edu} \\
  \And
  Tobias A. Kampmann \\
  Physics Department \\
  TU Dortmund University \\
  Dortmund 44227 Germany \\
  \texttt{tobias.kampmann@udo.edu} \\
  \And
  Jan Kierfeld \\
  Physics Department \\
  TU Dortmund University \\
  Dortmund 44227 Germany \\
  \texttt{jan.kierfeld@udo.edu} \\
}

\newcommand{\pvec}[1]{\vec{#1}\mkern2mu\vphantom{#1}'}

\newcommand{\comment}[1]{}

\usepackage{xcolor}

\begin{document}

\flushbottom
\maketitle


\begin{abstract}
Colloidal particles suspended in liquid crystals can
exhibit various effective anisotropic interactions that can be
tuned and utilized in self-assembly processes. 
We simulate a two-dimensional system of hard disks suspended in a
solution of dense hard needles as a model system for colloids suspended
in a nematic lyotropic liquid crystal.
The novel event-chain Monte Carlo technique
enables us  to 
directly  measure  colloidal interactions
in  a microscopic simulation with explicit liquid crystal particles
in the dense nematic phase.
We find a directional short-range attraction for disks
along the director, which triggers chaining parallel to
the director and 
seemingly contradicts the standard liquid crystal field theory result 
of a quadrupolar attraction with a preferred ${45\degree}$ angle. 
Our results can be explained by a short-range
density-dependent depletion interaction, which has been neglected
so far. Directionality and strength of the depletion interaction
are caused by the weak planar anchoring of hard rods.
The depletion attraction 
robustly dominates over the quadrupolar elastic attraction
if disks come close.
Self-assembly of many disks proceeds via intermediate chaining, which
  demonstrates that in lyotropic liquid crystal colloids
depletion interactions  play an important role
 in  structure formation processes. 
 \end{abstract}

\thispagestyle{empty}

\section*{Introduction}

Colloidal suspensions of nanometer to micrometer-sized particles in
a host fluid can form liquid and crystalline phases, but also
liquid crystalline mesophases if colloidal particles are, for example,
rod-shaped \cite{russel1990,Lu2013}.
Phase behavior and stability of a colloidal system can often be explained
based on effective interactions between colloidal particles
which arise from integrating out microscopic degrees of freedom
of the host fluid. The resulting
 effective interactions  govern colloidal stability, coagulation, 
flocculation, and structures that eventually self-assemble.

Colloidal mixtures containing different colloidal particles,
often of different size, 
feature additional effective interactions if degrees of freedom
of one species are integrated out \cite{louis2002}.
The most prominent example of additional effective interactions
in mixtures are depletion interactions, as
they arise, for example, in a mixture of large and small hard
spheres \cite{asakura1954,lekkerkerker2011}:
a short-range attractive depletion
interaction between large hard spheres emerges
because the excluded volume  for the small spheres
decreases (and, thus,
their available phase space increases) if
large spheres approach closer than one diameter of a small sphere.

Effective interactions become
even more interesting for colloids suspended
in anisotropic fluids such as liquid crystals (LCs), which can also be
seen as colloidal mixtures of larger colloidal particles suspended
in a liquid of small rod-like particles, in particular for lyotropic
LCs. Such LC colloids exhibit
anisotropic effective interactions between colloidal particles
if the LC is in an ordered, e.g., nematic phase \cite{Stark2001}.
The first studies of  LC colloids have been performed
  on latex spheres suspended in a lyotropic nematic LC
  (a micellar nematic phase of discoid type) \cite{Poulin1994,Raghunathan1996}.
The nematic LC phase forms typical defect-structures
  around a spherical inclusion depending on the anchoring
  conditions, ``Saturn-ring'' disclination rings  or a satellite hedgehog
  for normal anchoring and boojums for planar
  anchoring \cite{Terentjev1995,Kuksenok1996,Ruhwandl1997a}.

These defect-structures also induce long-range colloidal interactions
  between spherical inclusions.
  These 
  have been first explored systematically by Ramaswamy {\it et al.}
  \cite{Ramaswamy1996}, Ruhwandl and Terentjev \cite{ruhwandl1997}
    and by Poulin {\it et al.}
who also observed a chaining of water
droplets inside a LC \cite{poulin1997}.
The nature of these elastic LC-mediated
interactions strongly depends on the details of the interaction between
colloidal particle and LC host, i.e., how the
LC molecules or rods are anchored on the colloid
(normal, conic or planar, weak or
strong anchoring), and can be  of dipolar,
quadrupolar, or  even more complicated nature
\cite{ravnik2009, Pergamenshchik2010, smalyukh2018liquid}.
Such systems are promising candidates to realize a controlled
self-assembled structure formation by anisotropic interactions
\cite{musevic2006, skarabot2008,ognysta2008, lapointe2009,
  ognysta2011,tkalec2013}.
The elastic interactions in combination with different shapes
and sizes of the colloidal particles can be used to create a
multitude of different colloidal interactions
\cite{Liu2013,pergamenshchik2018,senyuk2019}.
LC colloidal  assemblies 
can be engineered, for example, by tuning the surface anchoring
or engineering nematic defect structures 
\cite{musevic2013,tkalec2013,musevic2017}.

Here, we consider a colloidal mixture of hard disks and hard needles,
which serves as a simple two-dimensional model system of
colloidal spheres suspended in a lyotropic LC.
{We note that  systems of had rods and hard disks in three dimensions are 
   qualitatively different as both disks and
   rods form LC phases in three dimensions
   \cite{Stroobants1984,Cuetos2008}.}
We are interested in the situation, where the hard needles are sufficiently
dense to form a nematic LC phase such
that they can mediate elastic  interactions between the
hard disks. 
This system is interesting from a theoretical point of view as
it can be interpreted, on the one hand, as a colloidal mixture, where
we expect depletion interactions between two hard disks embedded
into a fluid of shorter needles.
We also expect, on the other hand, to find
long-range elastic interactions mediated by the
nematic hard needle LC, at least for needles
shorter than the disks, such that a coarse-grained continuum
description is appropriate. This is the situation we want to
address in this paper. 
The interplay and competition of  (at least) two types
of effective interactions -- short-range depletion and long-range elastic --
has to be unraveled and will have
interesting consequences for the total effective interaction between
the disks.

Depletion interactions are well documented for the dilute isotropic
phase of needles
\cite{mao1995,yaman1998,koenderink1999,lin2001,chen2002,roth2003}
but much less is known
for a nematic host with the notable exceptions of Refs.\
\citenum{vanderschoot2000,Adams1998}, where
the limit of small spheres in long needles has been considered.
In the experimental study in Ref.\ \citenum{Adams1998} a chaining
of small spherical particles in a host of fd virus rods
parallel to the director has been found, which was attributed
to depletion attraction.

On the other hand,
the theory of colloidal LCs should apply with 
long-range elastic interactions mediated by director
field distortions in  the
nematic hard needles. Depending on
anchoring conditions and dimensionality of the system
dipolar interactions (falling off as $r^{-3}$ with the sphere separation
in three dimensions)
\cite{Lopatnikov1978,lubensky1998,fukuda2004}
or quadrupolar interactions (falling off as $r^{-5}$)
can occur \cite{Ramaswamy1996,ruhwandl1997,tasinkevych2002, mozaffari2011,
  tasinkevych2012}.
 Because hard needles
tend to align tangentially at a hard wall, we expect a quadrupolar
elastic interaction, which is characteristic for planar anchoring
at the colloidal disk
\cite{Ramaswamy1996,ruhwandl1997, mozaffari2011,tasinkevych2012,pueschel2016}
but also generic in two dimensions \cite{tasinkevych2002}. 
Both for dipolar and quadrupolar interactions the elastic interaction
is attractive and chaining of colloidal spheres
has been  experimentally
observed \cite{poulin1998,smalyukh2005,skarabot2008}.
Whereas dipolar interactions  prefer chaining of spheres
parallel to the director
axis in three dimensions \cite{lubensky1998,fukuda2004},
quadrupolar  interactions
prefer  an angle of approximately  ${30\degree}$ with respect
to the director axis  in three dimensions
\cite{mozaffari2011,pueschel2016} 
and a  ${45\degree}$ angle  in two dimensions \cite{tasinkevych2012}.
For our two-dimensional system of hard disks and hard needles
we thus expect the elastic interaction to  favor a ${45\degree}$ angle
if disks form chains.

Colloidal mixtures are also 
computationally challenging systems.
Effective interactions are essential to characterize stability and
potential self-assembly  into crystalline phases but hard to access
in a microscopic particle-based simulation. The process of integrating
out microscopic degrees of freedom corresponds to the numerical evaluation
of a potential of mean force between the colloidal species of interest.
In order to measure the potential of mean force all degrees
of freedom including, for example, relatively slow
large spheres in a  bath of small particles must be properly equilibrated.

The colloidal mixture of hard disks suspended in a nematic
host of hard needles
is particularly challenging as the hard needle system must be
fairly dense to establish a nematic phase.
While particle-based simulations exist for dilute rods in the isotropic phase
\cite{schmidt2001,kim2004},
the regime of a nematic host is fairly unexplored up to now and
simulations resorted to coarse-graining approaches \cite{pueschel2016}.
So far, only single inclusions \cite{Rahimi2017}  or confining
geometries \cite{Garlea2019} have been investigated by
particle-based simulations.
In order to calculate the effective interactions between hard disks
and their self-assembly 
we apply a novel Monte Carlo (MC) technique, the rejection-free
event-chain sampling technique.
The event-chain algorithm has been originally
proposed for pure hard sphere systems 
\cite{bernard2009}, and   recently generalized to
dense hard needle systems \cite{harland2017}, and is used here
to efficiently equilibrate the system, which allows us
to directly obtain the potential of mean force and, thus,
the effective interaction between
two hard spheres in a nematic hard needle host.

The simulation will reveal a surprising and robust tendency
for chaining of disks along the director
axis which seems to contradict the
chaining in a ${45\degree}$ angle with respect
to the director axis as predicted by
quadrupolar elastic interactions in two dimensions
\cite{tasinkevych2012}.  This turns out to be a result of a
dominant short-range depletion interaction, and
the analysis and explanation of this phenomenon
are the main topic of the present paper.

\section*{Results}

\begin{figure}
  \centering
  \includegraphics{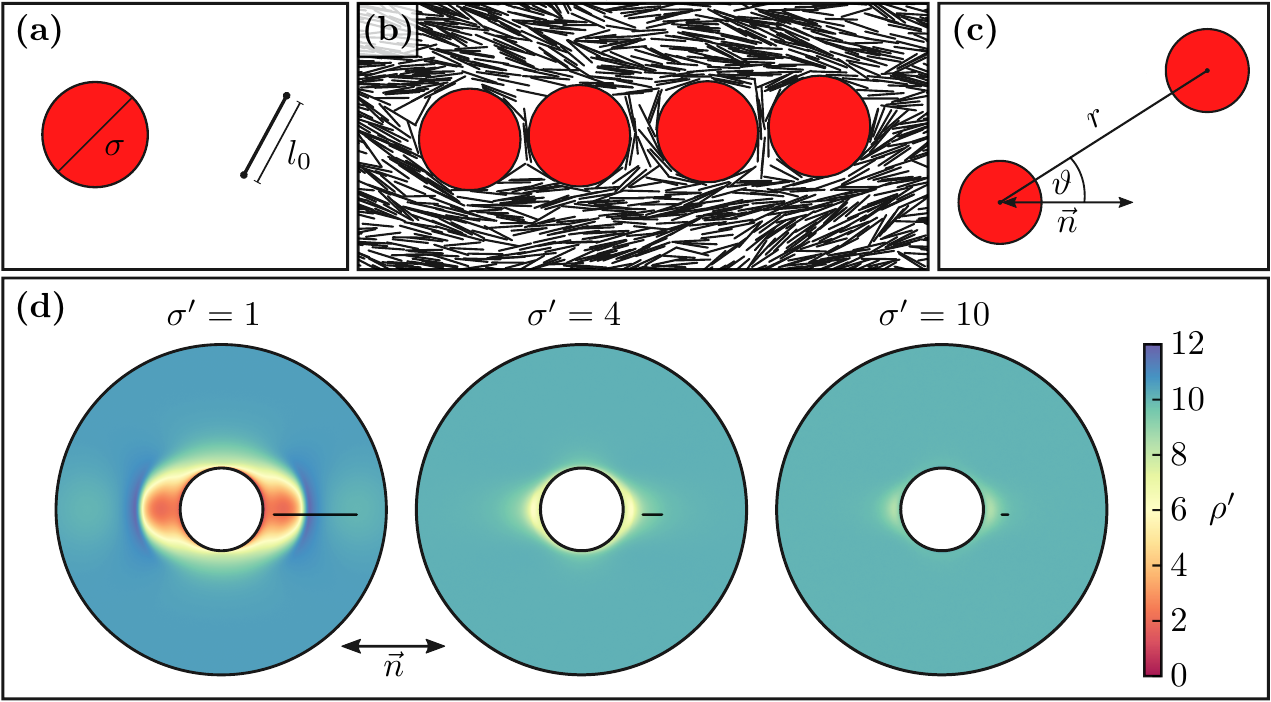}
  \caption{
    \textbf{(a)} Simulated particles:
    hard disks with diameter $\sigma$ and  hard
    needles with characteristic length $l_0$.
    \textbf{(b)} Simulation snapshot of disks ($\sigma/l_0=3$) forming chains
    along the director $\vec{n}$ (approximately in horizontal direction).
    On the outer disks and between disks surface depletion zones
    with low needle density form.
    \textbf{(c)} Disk distance $r$ and  disks angle  $\vartheta$ towards the
    director $\vartheta$ are used to describe the potential of mean force as
    well as other interactions.
    \textbf{(d)} Needle center of
    mass density distribution $\rho'(\vec{r})$ around a disk, relative to the
    director for different relative disk sizes $\sigma'=\sigma/l_0$
      (scale bars are one needle length $l_0$). 
    For small disks, there are distinct
    depletion zones in front of and behind a disk.
    Depletion zone extensions are $O(l_0)$.
    Overlap of these depletion zones gives rise to
    the density-dependent depletion interaction.
  }
  \label{fig_disk_needle_chaining}
\end{figure}

We use hard needles as a model system for a  lyotropic LC.
Hard needles can be viewed as two endpoints connected
by an infinitely thin, hard line,
see Fig.\ \ref{fig_disk_needle_chaining}(a),
and we sample in the MC simulation by moving the two endpoints. 
The needle length $l_0$ is used as unit length
for non-dimensionalization in the following (primed quantities are  
dimensionless).
In two dimensions needles can order 
at sufficiently high area densities
$\rho_\text{n}'\equiv \rho_\text{n} l_0^2 > 6$ in 
a Kosterlitz-Thouless transition into a quasi-ordered
nematic phase \cite{vink2014,harland2017}.
In the following, we focus on  a needle system
with $\rho_\text{n}'=  10$ well within this nematic regime.
We suspend hard disks of diameter $\sigma$ (${\sigma'\equiv \sigma/l_0}$)
as colloidal particles into the nematic hard needle system,
see Fig.\ \ref{fig_disk_needle_chaining}(b). 
The interaction between all particles, i.e., disk-disk,
disk-needle and needle-needle, is given by hard core potentials,
which  prohibit overlaps.
For fast sampling, 
we employ the rejection-free event-chain MC algorithm. 
More details on the simulation are provided in the
Methods section. 
The resulting needle-mediated
effective interactions between disks in this lyotropic system
is the focus of our investigation.

\subsection*{Chaining and depletion}

In the MC simulation, we observe a chaining of the disks along  the director
axis of the nematic needle phase, see
Fig.\ \ref{fig_disk_needle_chaining}(b). This chaining indicates a
strongly anisotropic directional 
attraction caused by the hard needles along the director axis.
The chaining axis coincides with two elongated zones depleted 
of  needles in the direction of the director on the surface of the disk,
as revealed by 
the needle density shown in  Fig.\ \ref{fig_disk_needle_chaining}(d).
Depletion zones have an extension of order $l_0$. 
Therefore, they are deep relative to the disk diameter and
very distinct for 
 small disks (${\sigma'=1}$).  For larger disks, the
depletion zones smear out  and their depth diminishes as compared
to the disks size.

\begin{figure}
  \centering
  \includegraphics{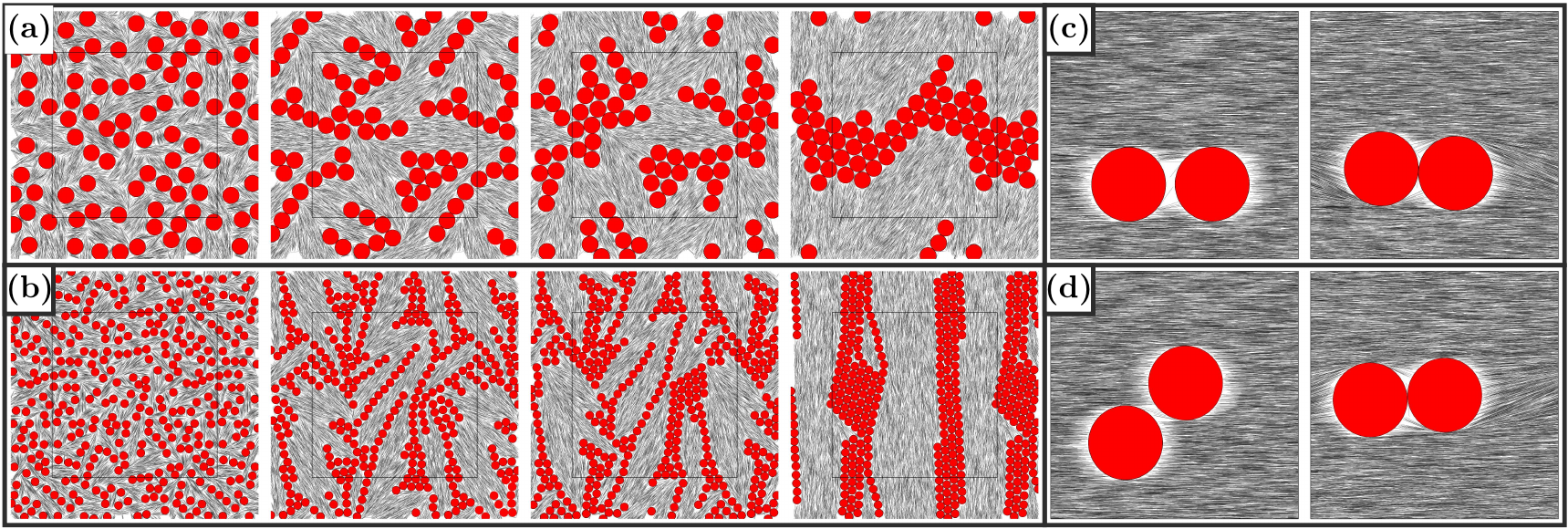}
  \caption{ \textbf{(a)} and  \textbf{(b)} Evolution of a multi-colloid system,
    containing  $40$ disks with diameter $\sigma = 2$ in
      \textbf{(a)}   $160$ disks
    with diameter $\sigma = 1$ in
      \textbf{(b)}. In both systems the same area is
    covered by disks, the needle density is $\rho'_\text{n} = 20$.
    To generate the initial configuration we place the disks
    overlap-free and add needles where they fit until the desired
    density is achieved.  The series of snapshots  shows the
    self-assembly of the disks first in chains and then into  hexagonal
    clusters. We show partly periodic images, extending
    the original system indicated by a black square.
    \textbf{(c)} and \textbf{(d)} Also in high
    density systems ($\rho'_\text{n}=100$)  disks align with the
    director.  \textbf{(c)} and \textbf{(d)} have different initial
     configurations, shown in the
     left column with disks at  $0\degree$ and $45\degree$ angles. Right column
     shows equilibrated state.
  }
  \label{fig:self-assembly}
\end{figure}

Chaining along the director axis also
governs self-assembly in
 simulations containing many disks shown
 in Fig.\ \ref{fig:self-assembly}(a) and (b).
 Here we observe self-assembly of hexagonal
disk crystals via  intermediate chain formation and subsequent
 aggregation of parallel chains.
The chaining along the director axis
 is very robust and occurs up to high needle densities
$\rho'(n)=100$, see Fig.\  \ref{fig:self-assembly}(c) and (d),
where the nematic LC becomes very stiff.

\subsection*{Effective disk-disk interaction}

\begin{figure}
  \centering
  \includegraphics{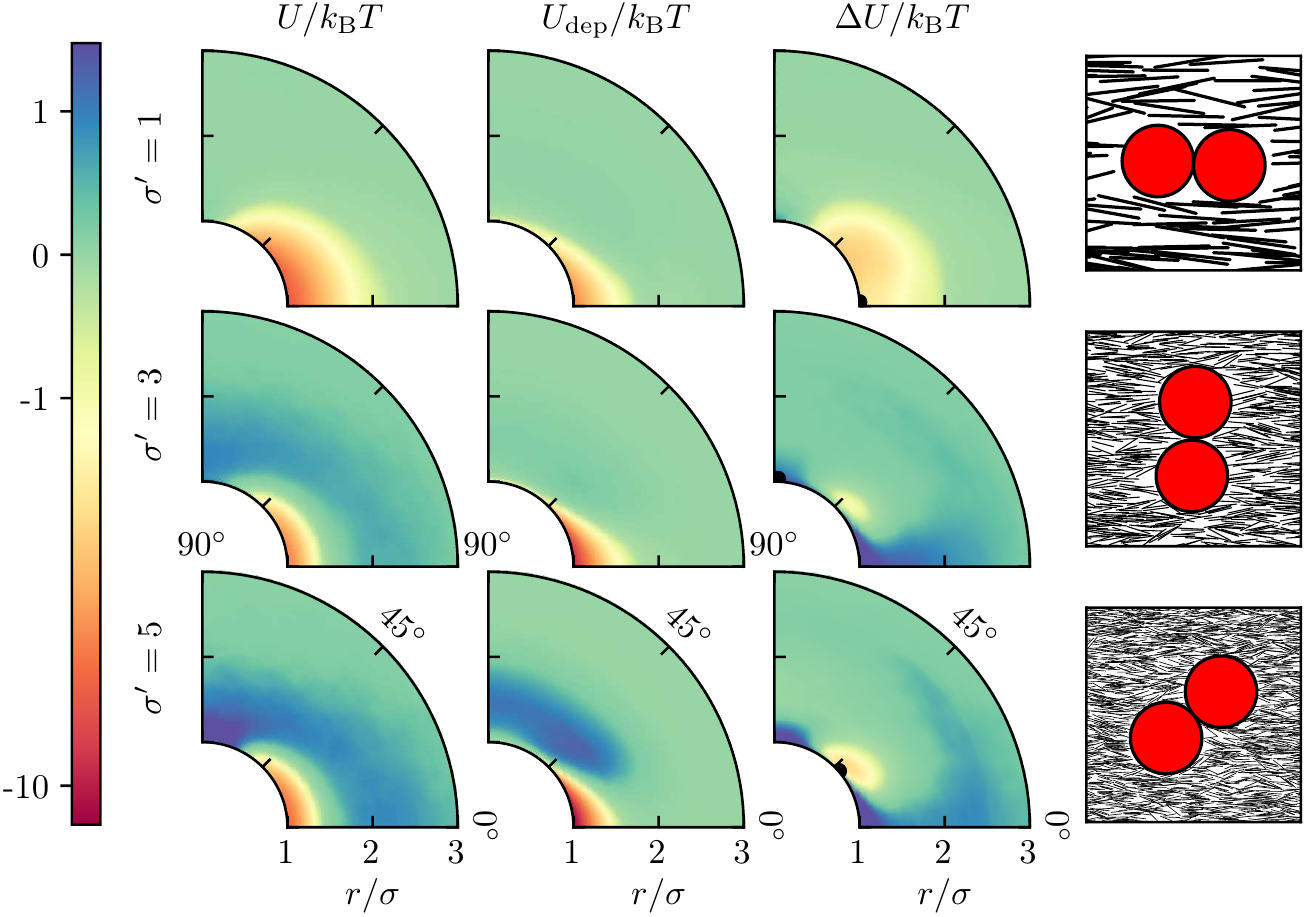}
  \caption{
    Contour plots of the measured interaction $U$, the calculated
    density-dependent depletion interaction $U_\text{dep}$
    and the difference of the two $\Delta U = U - U_\text{dep}$,
    showing the quadrupolar behavior.
    Columns represent the different interactions and rows describe different
    disk diameters $\sigma' = 1$, $3$ and $5$.
    The measurement was performed in a square system
    $L \times L$ with system size 
    $L=10 \sigma$ and needle density $\rho_\text{n}'=10$.
    On the right side are exemplary snapshots of configurations,
    marked with the black dot in the contour plot next to it.
  }
  \label{fig_potentials}
\end{figure}

To quantify the effective interaction between two hard disks suspended in
nematic hard needles, we use the potential of mean 
force ${U(r, \vartheta)\approx- k_\text{B} T \ln g(r, \vartheta)}$.
Here, $r$ is the distance between the disks and
$\vartheta$ the angle between the connecting line 
and the director $\vec{n}$, see Fig.\ \ref{fig_disk_needle_chaining}(c);
$g(r, \vartheta)$ is the pair distribution function and $k_\text{B} T$
the thermal energy. 
The event-chain MC simulation technique
is ideally suited for dense mixed
colloidal systems \cite{Kampmann2015,harland2017}
and allows us to obtain  
the positional and angular dependence of the
potential of mean force as shown   in 
Fig.\ \ref{fig_potentials}.

For small angles $\vartheta$ the interaction is attractive and
has its minimum  at the disk surface at an angle of
${\vartheta=0\degree}$. This explains the observed parallel
chaining of  disks.
For larger disks, the interaction also  develops a
repulsive part around ${\vartheta=90\degree}$.
The range of the interaction is decreasing relative to the 
disk size and of order of $l_0$. 

The effective interaction can  be described as the sum of
two interactions, which are 
the depletion interaction of short range $l_0$
resulting from overlapping depletion zones around a hard
disk, and a power-law
quadrupolar elastic interaction resulting from the elastic energy
of nematic hard needles.
 Because hard needles
 tend to align tangentially at a hard wall, we expect a quadrupolar
 elastic interaction falling off as $r^{-4}$ and with a
 $\cos(4\vartheta)$ angular characteristic
 \cite{tasinkevych2002, dolganov2006,tasinkevych2012}.
 Therefore,  the elastic interaction is maximally
 attractive for  ${\vartheta=45\degree}$ and repulsive
 both for ${\vartheta=90\degree}$ and ${\vartheta=0\degree}$.
The  repulsive interaction at ${\vartheta=90\degree}$
can thus be explained by dominating quadrupolar
elastic interactions, which
becomes more relevant for larger disks.  The interaction minimum at
${\vartheta=0\degree}$, however, comes as a surprise in view of the
quadrupolar interaction of standard LC field theory.
This minimum can only be explained by the dominance of the   attractive
depletion interaction
induced by the nematic hard needles, which must be directed
along the depletion zones parallel to the director.

\subsection*{Density-dependent depletion interaction}

In order to calculate the depletion interaction caused by the 
 complex anisotropic and smeared shapes of the depletion zones in Fig.\
\ref{fig_disk_needle_chaining}(d),
we use a density-dependent description of
depletion interactions and find 
\begin{align}
  U_\text{dep}(\vec{r})
  &=  -  k_\text{B} T \rho_\text{n} \int
   \left(1 - \frac{\rho(\pvec{r})}{\rho_\text{n}}  \right)
  \left(1 - \frac{\rho(\pvec{r}-\vec{r})}{\rho_\text{n}}\right)
  \mathrm d \pvec{r}
  \, ,
  \label{eq:Udep}
\end{align}
where $\rho(\vec{r})$ is the local needle density and $\rho_\text{n}$ the
average needle density. The depletion interaction is proportional
to a generalized overlap area of depletion regions of two disks at distance
$\vec{r}$, where the
depletion region of a disk at $\pvec{r}=0$ is given by the region with 
$1 - {\rho(\pvec{r})}/{\rho_\text{n}}\approx 1$, which can correctly capture
the  complex shaped depletion zones in Fig.\
\ref{fig_disk_needle_chaining}(d).
A  detailed derivation of eq.\ (\ref{eq:Udep}) is presented
in the Methods section.

The 
depletion zones are mainly  shaped by the elastic interactions
in the nematic phase 
via anchoring conditions at the disk surface.
Hard rods exhibit planar anchoring at a hard disk.
In the simulation the anchoring is only parallel on average, and
fluctuations weaken the anchoring considerably. 
We only  find a weak entropic planar anchoring (quantified below)
such that
the needle orientation only deviates little from the director
orientation at the disk surface. This explains 
the  elongated  depletion zones of extension of order $l_0$
(see Fig.\ \ref{fig_disk_needle_chaining}(d)).
This also results
in an overlap area $\sim (\sigma l_0^3)^{1/2}$
from standard geometric arguments \cite{lekkerkerker2011}, such that
$U_\text{dep} \sim k_BT \rho_\text{n}' \sigma'^{1/2} $ is
expected.
Moreover, this gives rise 
to  a strongly anisotropic  depletion attraction, which is 
maximal parallel to the director ($\vartheta=0\degree$). 
We can calculate the 
depletion  interaction numerically using measured density
profiles from simulations and (\ref{eq:Udep}); the result is
shown in Fig.\ \ref{fig_potentials}.

\subsection*{Revealing  the residual   elastic interaction}

\begin{figure}
  \centering
             \includegraphics{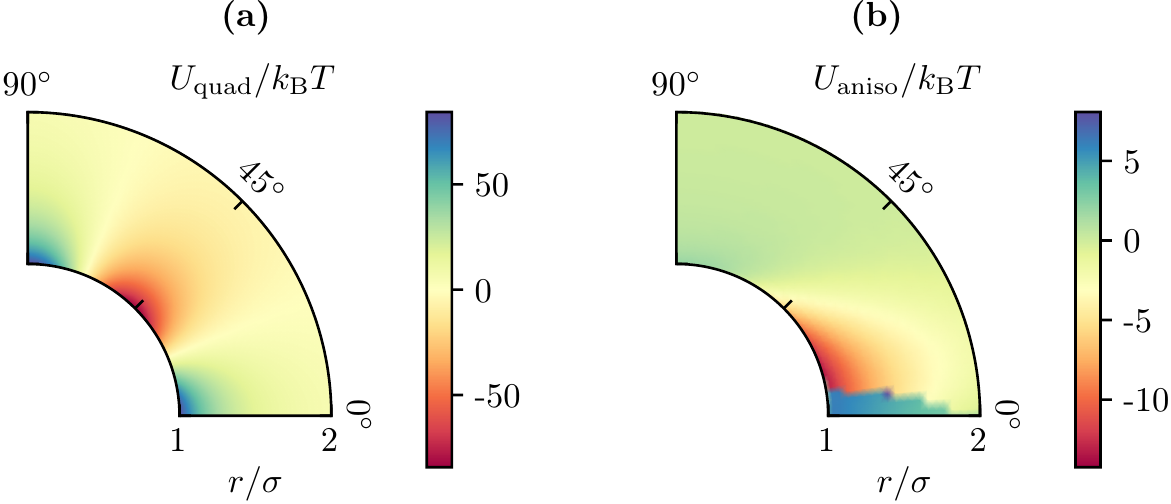}
  \caption{
    \textbf{(a)}
    Quadrupolar interaction  (\ref{eq:quadrupolar_energy})
    of two disks in a liquid crystal with
    planar boundary condition on the disk's surface (with
    ${K / \rho_\text{n}'^2 k_\text{B}T = 0.358}$, see text).
    This
    is a field theoretical result for the far-field behavior,
    assuming weak distortions of the director and using a one-constant
    approximation.
     \textbf{(b)}
     Numerically calculated quadrupolar interaction between two disks
    in a system of size
  $L=10\sigma$ in the presence of elastic anisotropy  and weak anchoring
  ($K_1/K_3=0.1$ and $w=0.5$, see (\ref{eq:wfit}), and
  $K_3 / \rho_\text{n}'^2 k_\text{B}T = 0.63$).
  }
  \label{fig_quadrupolar_interaction}
\end{figure}

In order to uncover additional effective interactions apart
from  depletion, we subtract the numerically calculated
depletion  interaction from the
measured total potential of mean  force and obtain
the residual interaction 
${\Delta U = U - U_\text{dep}}$. All three interactions are  shown 
in Fig.\ \ref{fig_potentials}.
This reveals the presence of another interaction, which can
actually be identified as the elastic interaction from 
LC field theory.
Hard needles
tend to align tangentially at a hard wall, such that
we have planar boundary conditions.
For two disks suspended in a nematic LC
with planar boundary conditions, LC field theory
predicts 
\cite{tasinkevych2002, dolganov2006}
\begin{align}
  U_\text{quad}(r, \vartheta)
  &\approx \frac{3 \pi K \sigma^4}{4}
  \frac{\cos(4 \vartheta)}{r^4}
  \, .
  \label{eq:quadrupolar_energy}
\end{align}
This result is based on the  one-constant approximation, i.e.,
assuming an isotropic elasticity of the nematic LC
with a single elastic constant $K$. This elastic
constant can be calculated  for hard needles
in two dimensions as 
${K / \rho_\text{n}'^2 k_\text{B}T = 0.358}$
by adapting a method from
Straley \cite{straley1973}.
The quadrupolar interaction (\ref{eq:quadrupolar_energy}) is shown
in Fig.\ \ref{fig_quadrupolar_interaction}(a) and
explained in more detail in the Methods section.

The field-theoretical quadrupole interaction
(\ref{eq:quadrupolar_energy}) is
qualitatively similar to the residual interaction $\Delta U$ from
Fig.\ \ref{fig_potentials}  but exhibits characteristic
deviations.  The main differences are:
(i) a much weaker interaction strength, (ii)
a distorted quadrupolar symmetry,  and (iii) 
for small disks (${\sigma' = 1}$), a missing repulsive regime around 
${\vartheta = 0\degree}$ and ${\vartheta = 180\degree}$.
The low interaction strength of the measured interaction (i)
is caused by the  weak parallel anchoring of hard rods
at the disk surface, which we find in the simulation (and which
we quantify below). 
In the theoretical result (\ref{eq:quadrupolar_energy}), on the other hand,
strict planar boundary conditions, i.e., infinitely 
strong parallel anchoring is assumed.
The distorted quadrupolar symmetry (ii) becomes manifest
in  differently strong
repulsion zones for ${\vartheta=0\degree}$ and ${\vartheta=90\degree}$.  This
is caused by the elastic anisotropy of the nematic hard needle phase, i.e.,
different elastic constants ${K_1 \neq K_3}$ for orientation fluctuations
perpendicular ($K_1$) or parallel ($K_3$) to the director, which
are not captured in the  one-constant
approximation (${K = K_1 = K_3}$) employed in
(\ref{eq:quadrupolar_energy}).
Adapting  the  method from Ref.\  \citenum{straley1973}, we find
$K_1/K_3 \approx 0.14$ for hard needles in two dimensions (our
above value ${K / \rho_\text{n}'^2 k_\text{B}T = 0.358}$ is actually
defined as  $K=(K_1+ K_3)/2$.
From  fitting   mean nematic orientation field around a disk we confirm
a value of $K_1/K_3 \approx 0.1$ below. 
The weak repulsion for small disks
(${\sigma'=1}$) at ${\vartheta=0\degree}$ and  ${\vartheta=180\degree}$ (iii)
can be explained by the pronounced  density
dependence of the elastic constant ($K\propto \rho_\text{n}^2$)
via a correlation effect with the
depletion interaction.  Since the depletion zones are of very low
density and pronounced in this direction, especially for small disks,
they also weaken the elastic interaction in
this direction.

This interpretation is supported by a numerical calculation
 of the elastic interaction between two disks
in the presence of elastic anisotropy $K_1/K_3 = 0.1$ and
weak parallel anchoring (as quantified below), which is shown
in Fig.\ \ref{fig_quadrupolar_interaction}(b). Details of the numerical 
calculation are explained in the Methods section.
The numerically calculated elastic interaction only misses
the correlation effect (iii) and, indeed,  resembles
the residual interaction $\Delta U$ from
Fig.\ \ref{fig_potentials} more closely in its distorted shape.

Both depletion and  quadrupolar interaction are
attractive and promote chaining of disks, however, at
angle $0\degree$ if depletion dominates and around
$45\degree$ if the elastic quadrupolar interaction dominates.
We see a robust $0\degree$-chaining which points at
a dominating depletion interaction. 
In order to explain this dominance  we need to 
quantify the weak anchoring for the hard needle  and disk system,
which is central both for directing the 
depletion interaction in  $0\degree$ direction as well
as for weakening the residual elastic interaction.

\subsection*{Weak anchoring strength and elastic anisotropy in
 quadrupolar interaction}

Both anchoring strength and elastic anisotropy  can be quantitatively
analyzed by investigating the needle orientation field $\Phi(\vec{r})$
around a single disk, which is suspended
in a nematic phase with $\Phi =0$ asymptotically at the system boundary
($\Phi(\vec{r})$ is the angle
of the local director field $\vec{n}(\vec{r})$
with this asymptotic orientation, i.e., $\vec{n}=(\cos\Phi, \sin\Phi)$;
$\Phi$ is  defined with respect to the asymptotic
director orientation, which determines the $x$-axis). 
By comparing numerical MC results for hard needles and field theory
calculations we can deduce both anchoring strength and elastic anisotropy.

The free energy of a two-dimensional
LC with anchoring on a surface is given by
the Frank-Oseen elastic energy \cite{Straley1971} and a surface
anchoring potential:
\begin{align}
  F &= \int \mathrm d A \left[
  \frac{K_3}{2}  (\partial_x \Phi)^2
  +\frac{K_1}{2}  (\partial_y \Phi)^2 \right] 
  +\frac{W}{2} \int \sin^2(\Phi_0 - \Phi) \mathrm d l \, .
  \label{eq:F}
\end{align}
The first integral is the approximate
Frank-Oseen free energy $F^\text{2D}_\text{FO}$
with elastic constants
$K_1$ and $K_3$ (see Methods section)
and the second integral over 
the disk's surface represents the anchoring energy with anchoring strength $W$.
The anchoring depends on the boundary condition $\Phi_0$ at the disks surface, 
which is parallel anchoring for hard needles at a hard disk
(i.e., $\Phi_0 = \theta-\pi/2$).
The numerical minimization of this free energy is discussed
in the Methods section. 
Introducing dimensionless quantities (marked with a tilde)
by measuring the free energy in units  of the elastic constant $K_3$ and
distances in units of the  disk radius $R =  \sigma/2$,
we find two control parameters,
the elastic anisotropy $K_1/K_3$ and 
the relative anchoring strength $w\equiv  W\sigma/2K_3$.

We fit  the free energy minimization results to
MC simulation data  for the mean orientation field
$\Phi(\vec{r})$ of the needles (see Methods section).
Fitting $\Phi(\vec{r})$ in the range
$2r/\sigma =1.5-3$ and doing so for nematic densities $\rho_\text{n}'=10-20$
and disk sizes $\sigma'=1-10$ in systems of size
$L=6\sigma$ we obtain the best fit for
\begin{align}
  \frac{K_1}{K_3} &\approx 0.1 ~~\mbox{and} ~~~w\approx 0.45-0.50.
  \label{eq:wfit}
\end{align}
In principle, the relative anchoring strength $w$ can
depend on $\rho_\text{n}'$ and $\sigma'$.
We only find a very weak decrease from $w=0.5$ to $0.45$
with density for
$\rho_\text{n}'=10-20$.
The result for the elastic anisotropy is in good agreement with
our above finding $K_1/K_3 \approx 0.14$ using the method
from Ref.\  \citenum{straley1973}.
Both  results (\ref{eq:wfit})
are also very similar to results from Ref.\ \citenum{Galanis2010}
   for granular rods in two dimensions obtained in a cavity geometry. 
This confirms that the hard needle nematic phase has indeed
a pronounced elastic anisotropy and only a weak effective
anchoring strength at the hard disk surfaces;
this anchoring is purely entropic as reflected by 
$w={\rm const}$ implying
$W/\sigma \sim K  \sim k_BT \rho_\text{n}'^2$.

\subsection*{Depletion dominates  quadrupolar interaction
   for chaining along director}

  In the chaining of disks  the
  elastic quadrupolar interaction competes with the depletion interaction.
  For weak anchoring the quadrupolar interaction is $\propto w^2$
  \cite{tasinkevych2012}. From eq.\ (\ref{eq:quadrupolar_energy})
  we expect a quadrupolar interaction strength
 $U_\text{quad}/{k_BT} \sim 
 w^2 \rho_\text{n}'^2$  at the disk surface.
 This competes with a depletion interaction of strength
 $U_\text{dep}/k_BT  \sim  \rho_\text{n}' \sigma'^{1/2} $.
  For weak anchoring with 
 $w^2\rho_\text{n}'\sigma'^{-1/2} \ll 1$ 
 the depletion interaction dominates as observed in our simulations.
 This explains the robustly observed $0\degree$-chaining
 along the director.
 At high densities, the quadrupolar interaction becomes
 more relevant. 
 We performed simulations up to very high needle densities
 $\rho_\text{n}'=100$ and still observe robust   $0\degree$-chaining
 as shown in Fig.\ \ref{fig:self-assembly}.

\section*{Discussion}

We modeled lyotropic LC colloids as hard disks in a suspension of hard
needles in a two-dimensional system.
Simulations with an 
event-chain MC algorithm showed a chaining of disks along  the director
axis of the nematic needle phase.
This chaining
is caused by a depletion interaction due to elongated depletion
zones behind and in front of the disks parallel to the director.
This depletion interaction is only accessible in
efficient microscopic simulations with explicit rods.
Elongated depletion zones are caused by the
weak planar  anchoring of hard needles  at hard disks. 
 Calculating the depletion interaction with a
 density-dependent depletion theory
 and subtracting the depletion interaction we
 reveal  a residual elastic quadrupolar interaction which is
 mediated by the  director distortions around the  disks.
 A quadrupolar elastic interaction is consistent with
 the planar anchoring.
 The elastic interaction is weakened because hard needles are only
 weakly anchored  and the angular dependence
 is characteristically deformed because  of a pronounced
 elastic anisotropy of the nematic needle phase.

 Both depletion and quadrupolar interaction are attractive
 and can give rise to chaining, in principle, but
 the depletion favors a $0\degree$-angle with the director axis
 while the quadrupolar interaction would favor $45\degree$. 
 In our simulations  for densities up to  $\rho_\text{n}' = 100$ and
 disk sizes up to ${\sigma' = 10}$ we only find $0\degree$-angle
 chaining indicating a robustly dominating depletion, which is
 mainly due to the weak planar anchoring. 
  This type of chaining
 also  governs the intermediate states in self-assembly
 of many colloidal disks into crystals as shown in
 simulation in Fig.\  \ref{fig:self-assembly}.

A natural continuation of our work will be the investigation
of  hard spheres suspended in hard rods (such as spherocylinders)
in three spatial dimensions. 
We believe that the two-dimensional simulation can
already provide results,
which qualitatively apply  also to the
the three-dimensional case, while giving a
significant simulation speedup.

In three spatial dimensions, 
experiments \cite{poulin1998,smalyukh2005}
and numerical  field theory
\cite{mozaffari2011,pueschel2016} found
quadrupolar interactions with 
an interaction minimum and chaining of colloidal spheres
at a $30\degree$-angle for planar anchoring, which
corresponds to a $45\degree$ in two dimensions \cite{tasinkevych2012}.
The field-theoretic continuum approaches did not capture
depletion interactions, however, while experiments in Refs.\
\citenum{poulin1998,smalyukh2005} were
not dealing with lyotropic systems. 

With respect to applications, 
our results show that, in lyotropic LC colloids,
depletion interactions can play an important role
in structure formation. By exploiting their dependence
on particle shape they could provide an 
 additional tool to control the 
 structure formation process.
 { Assuming that our result  carry over to three-dimensional
   systems, we expect that the range of the 
   depletion interaction mediated by the lyotropic LC
   is given by the rod length $l_0$.
   In order to develop similar observable consequences for chaining
   as in the simulations, $l_0$ should not be orders of magnitude
   smaller than colloidal spheres; this requires
     fairly long lyotropic rod-like particles.
   The depletion interaction
 dominates for weak planar anchoring as realized by hard rods and
 hard spheres. Experimental investigations for these types of systems are
still rare at present.}
An ideal system to test the predictions experimentally
are fd virus rods. 
In the experimental study in Ref.\ \citenum{Adams1998}
very small hard spherical particles {($\sim 20{\rm nm}$)}
in a host of hard fd virus rods {(length $\sim 900{\rm nm}$)}
in the nematic phase 
have been studied and, indeed, an ordering 
parallel to the director has been found, which was attributed
to depletion attraction and is in line with our observations. 
{Our results suggest that future experiments with micron-sized
  colloidal particles suspended in a nematic  fd virus phase
  could display dominant depletion interaction effects including
  colloidal self-assembly via parallel chaining.}

\section*{Methods}

\subsection*{Simulation method}
\label{sec_simulation}

\subsubsection*{Event-chain Monte-Carlo}

To simulate hard disks in a suspension of hard needles the event-chain MC
algorithm for many-body interactions is used \cite{bernard2009, michel2014,
  harland2017}.
  This is a rejection-free MC method that 
 performs very well for dense systems \cite{Kampmann2015}, which
 makes it an excellent choice for needle systems in the nematic phase,
 for which it has been adapted in Ref.\ \citenum{harland2017}. 
 The event-chain MC is a Markov-chain MC scheme, which fulfills
 maximal (rejection-free) global balance 
 rather than detailed balance. Global balance
 is achieved by introducing lifting moves \cite{michel2014}.
For hard spheres or needles a lifting move is the transfer of
 MC displacement from one particle to another particle. 
This means in a MC move only one particle at a
time is moved along a line until it contacts another object.
For the application to hard needles, one of the two endpoints are
moved and, upon collision with another
needle, the remaining MC move distance is
transferred to one of the endpoints of this needle. To which of the
endpoints it is transferred
depends on the collision point along the needle
\cite{harland2017}. In the presence of additional disks, MC displacement
is also transferred to disks if a needle collides with the disk
and vice versa.
Collision detection is the computational bottleneck, and we
use a sophisticated  neighbor list system to achieve
high simulation speeds also in the nematic phase (see Supplementary 
Information).

\subsubsection*{Hard needle liquid crystal}

We model the molecules of a lyotropic LC  as hard needles, which
consist of two endpoints connected by an infinitely thin, hard line (see
Fig.\ \ref{fig_disk_needle_chaining}(a)).
For efficient sampling, the distance of the two endpoints, i.e. the
length $l$ of the hard needle can fluctuate around its
characteristic length $l_0$  in order to allow
for independent motion of both endpoints in the MC simulation;
the needle length $l$  is restricted by a hard potential $V_\text{n}(l)$
with  $V_\text{n}(l) = 0$ for  $l/l_0 \in [0.9, 1.1]$ and
infinite $V_\text{n}(l)$ else,  
such that $l_0$ is the effective needle length.
Restricting the needle length is also necessary for our neighbor lists, which
accelerate the simulation.

The interaction potential
$V_\text{nn}$ between two
hard needles is either infinite when they overlap or zero else.
This results in a lyotropic transition from isotropic to nematic
upon increasing the needle density $\rho_\text{n}$.
In a system with $N_\text{n}$ needles
the density $\rho_\text{n} = N_\text{n}/A_\text{free}$ is
defined using the available 
free area $A_\text{free}$ (reduced by the area occupied by disks).
The
simulation box is a square with edge length $L$ and 
periodic boundary conditions.
{In a typical system of size $A_\text{free} = 900 l_0^2$
  we simulate $N_\text{n}= 9000$ needles
  at a needle density $\rho_\text{n}'=10$. 
}

Liquid crystalline ordering is described by the standard
tensorial order parameter 
\begin{align*}
  Q_{\alpha \beta}
  = 
  \frac{1}{N_\text{n}} \sum_{i=1}^{N_\text{n}}
  \left(
    \vec{\nu}^i_\alpha 
    \vec{\nu}^i_\beta 
    - 
    \frac{\delta_{\alpha \beta}}{2}
  \right) 
  \, ,
\end{align*}
which is invariant under needle inversion; 
 $\vec{\nu}^i$ is the orientation of  needle $i$.

\begin{figure}
  \centering
  \includegraphics{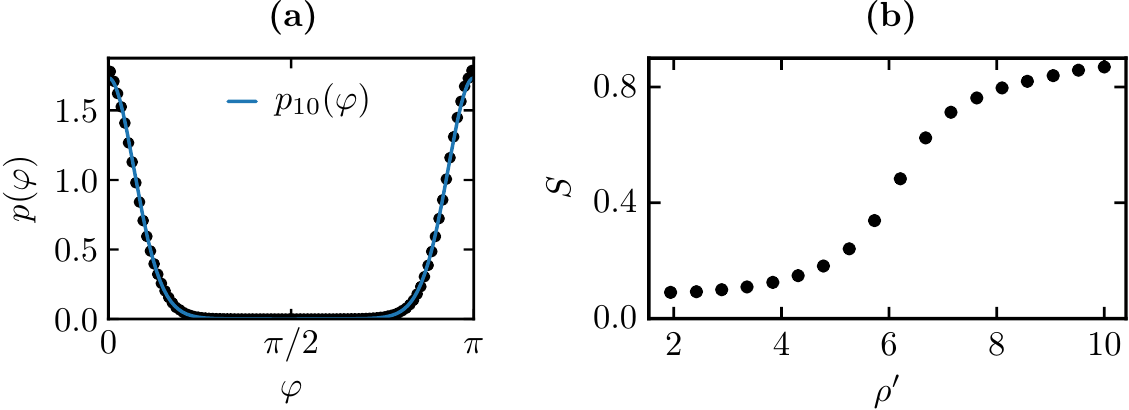}
  \caption{
    \textbf{(a)}
   Measurement of the probability distribution of angles 
   $\varphi$ with the director (black dots).
   The probability function $p_{10}(\varphi)\sim \exp(10 \cos^2\varphi)$
   (blue line) matches the MC data for  $\rho_\text{n}'=10$.
    \textbf{(b)}
   Scalar order parameter $S$ as a function of the needle density
   ${\rho_\text{n}' = \rho l_0^2 = N_\text{n}
     l_0^2/A_\text{free}}$.  The transition
   from the isotropic to the nematic phase occurs roughly at
   ${\rho_\text{n}' \approx 6}$ and is of Kosterlitz-Thouless type.
   We consider  needle systems with densities
   ${\rho_\text{n}' \ge 10}$ as in the nematic phase and use
   $\rho_\text{n}' =10$ for the potential of mean force calculation
    and local
    needle density measurements.
  }
  \label{fig_order}
\end{figure}

The scalar order parameter ${S \in [0, 1]}$ is calculated by diagonalizing the
matrix, where ${\lambda=2S}$ is the biggest eigenvalue.  The corresponding
eigenvector is the director $\vec{n}$.  The scalar order parameter $S$
measures the degree of alignment in the system and  is zero in a perfectly
isotropic phase and unity in a perfectly aligned
nematic phase. Figure \ref{fig_order} shows the order parameter
$S$ as a function of density for the two-dimensional
hard needle system (see also Ref.\ \citenum{harland2017});
the system quasi-orders above ${\rho_\text{n}' \approx 6}$. 
In our two-dimensional system, this transition  is a Kosterlitz-Thouless
transition into a quasi-ordered nematic state \cite{vink2014}.

\subsection*{Elastic energy, weak anchoring  and elastic anisotropy}

For a nematic LC the state of perfect alignment is
the energetically preferred configuration.
But this state is almost never reached because of  thermal fluctuations,
boundary conditions, or topological defects.  Often effective free energies
from field theories are used to describe deviations from homogeneous
alignment.  For a director field $\vec{n}(\vec{r})$ in three
dimensions, the energy of
small distortions can be captured with the Frank-Oseen free
energy $F_\text{FO}$ \cite{frank1958}
with three elastic constants $K_i$, which determine the
energy cost of the three different distortions  splay
($K_1$), twist ($K_2$) and bend ($K_3$).
In a two-dimensional system, twist distortions are not possible and, therefore,
${K_2=0}$ resulting in \cite{Straley1971}
\begin{align}
  F^\text{2D}_\text{FO} 
  &=  \int \mathrm d A \left[
  \frac{K_1}{2} 
  (\nabla \vec{n})^2 +
   \frac{K_3}{2}  (\nabla \times \vec{n})^2 \right]
  \approx  \int \mathrm d A \left[
  \frac{K_3}{2}  (\partial_x \Phi)^2
  +\frac{K_1}{2}  (\partial_y \Phi)^2 \right]
  \label{eq_frank_oseen_energy2D}
\end{align}
with the needle orientation field $\Phi(\vec{r})$
related to the director via  $\vec{n}=(\cos\Phi, \sin\Phi)$.
We assume  $\Phi =0$ corresponding to $\vec{n}||\vec{e}_x$
asymptotically at the system boundary.
The last approximation is the leading order in
an expansion in deviations from $\Phi=0$ and neglects
non-linear terms in $\Phi$ \cite{Galanis2010}.

The elastic constants $K_1$ and $K_3$ can be
calculated by adapting a  method described by
Straley \cite{straley1973} to two dimensions.
Strictly speaking the pronounced logarithmic orientational
fluctuations in two dimensions will give rise to a
renormalization at large scales \cite{Nelson1977}, which we ignore in
our finite size system.
We obtain 
${K_1 / \rho_\text{n}'^2 k_\text{B}T = 0.086}$ and
${K_3 / \rho_\text{n}'^2 k_\text{B}T = 0.63}$ for the nematic phase
with $\rho_\text{n}'=10$.
This result is based on  the angular distribution
$p_A(\theta) \propto \exp(A \cos^2\theta)$
of needle orientations $\vec{\nu}=(\cos\theta, \sin\theta)$ with
$A=10$ matching  our  MC simulations 
in the nematic phase with  $\rho_\text{n}'=10$ (see Fig.\ \ref{fig_order}).
In the one-constant approximation,
we assume $K=\frac{1}{2}(K_1+ K_3)$ resulting in
${K / \rho_\text{n}'^2 k_\text{B}T = 0.358}$; for this value
of $K$  the quadrupolar interaction (\ref{eq:quadrupolar_energy})
is shown in 
\ref{fig_quadrupolar_interaction}.

In order to describe colloidal disks in a nematic needle phase faithfully,
we need too take into account  elastic
anisotropy $K_1 \neq K_3$ and
a finite  parallel anchoring with an anchoring strength $W$.
In order to quantify the elastic anisotropy  $K_1/K_3$ and
the relative anchoring strength $w=W\sigma/2K_3$, we
consider the needle orientation field $\Phi(\vec{r})$
around a single disk, which is suspended
in a nematic phase with $\Phi =0$ asymptotically at the system boundary
and quantitatively compare MC simulation data
with minimization of the  free energy (\ref{eq:F})
containing both the anisotropic 2D Frank-Oseen elastic energy and the
anchoring potential. 
Free energy minimization  using  the linearized approximation in
the Frank-Oseen part 
(\ref{eq_frank_oseen_energy2D})
results in the  partial differential equation
\begin{align}
  \left(\partial_{\tilde{x}}^2
    + \frac{K_1}{K_3} \partial_{\tilde{y}}^2 \right) \Phi
  &=0~~(\tilde{r}>1),~~~~
      \partial_{\tilde{x}} \Phi + \frac{K_1}{K_3} \partial_{\tilde{y}} \Phi
           =  \frac{w}{2} \sin\left(2(\Phi -\Phi_0)\right)
    ~~(\tilde{r}=1).
    \label{eq:PDE}
\end{align}
in dimensionless units $\tilde{\vec{r}} \equiv 2\vec{r}/\sigma$. 
Analytical solutions are only available in the
one-constant approximation $K_1 = K_3$, where \cite{burylov1994}
\begin{align}
  \Phi(r, \vartheta) &= 
    - \arctan
    \left[
      \frac{(\sigma/2r)^2 p(w) \sin(2\vartheta)}
            {1 - (\sigma/2r)^2 p(w) \cos(2\vartheta)}
    \right]
    \qquad \text{with} \qquad 
  p(w) = \frac{2}{w}\left( \sqrt{1 + \frac{w^2}{4}} - 1 \right)
  \label{eq:burylov}
\end{align}
(both $\Phi$ and $\vartheta$ are defined with respect to the asymptotic
director orientation, which determines the $x$-axis). 
For the full problem~\eqref{eq:PDE}  we have to resort to numerical
solutions 
by finite element methods (using MATHEMATICA).

\begin{figure}
  \centering
  \includegraphics[width=0.99\linewidth]{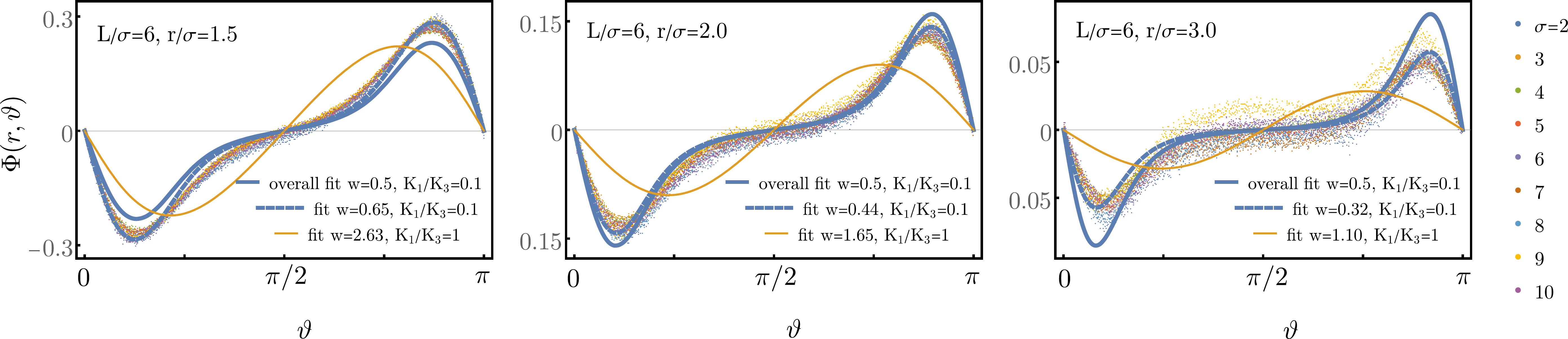}
  \caption{
    MC simulation results for  $\Phi(r,\vartheta)$ as a function
    of $\vartheta$ and for $r/\sigma=1.5,2.0,3.0$
    for parameters  $\rho_\text{n}'= 10$,
    $\sigma=2,...,10$, and $L=6\sigma$.
    Data for different $\sigma$ collapse
    fairly well. Blue dashed lines: Least square fits of data for each
    $r/\sigma$ with numerical solution
    of the PDE (\ref{eq:PDE}) in the same geometry as simulation,
     for $K_1/K_3=0.1$, 
    and with $w$ as fit parameter.  Blue solid lines:
    Analogous least square fit of data for all
    $r/\sigma=1.5-3.0$. Yellow lines:
     Least square fits with one-constant result (\ref{eq:burylov})
      ($K_1/K_3=1$)  with $w$ as fit parameter.
  }
  \label{fig:plotall}
\end{figure}

In Fig.\ \ref{fig:plotall} we plot
MC simulation data for the director orientation field
$\Phi(r,\vartheta)$ as a function
of $\vartheta$ (i.e., along circles) for different rescaled $r/\sigma$
and for different system sizes such that $L/\sigma=6$ remains fixed.
Rescaling of lengths with $\sigma$ 
results in good data collapse for different $\sigma$.
The characteristic shoulder of $\Phi(r,\vartheta)$ as a function of
$\vartheta$ can only be explained by an elastic anisotropy
$K_1/K_3\approx 0.1$.
Fits with eq.\ (\ref{eq:burylov}) from elastically isotropic
one-constant approximation
($K_1=K_3$) remain unsatisfactory. 
We perform fits for specific $r/\sigma$ and overall fits within the
whole range $r/\sigma=1.5-3.0$ with the numerical solution
of the PDE (\ref{eq:PDE}) in the same geometry as the MC simulation
(square box with $L/\sigma=6$ and director oriented
along the diagonal as fixed by Dirichlet boundary conditions)
using  $K_1/K_3=0.1$
  and  $w$ as fit parameter. 
  The fit results for $w$ are weakly $r$-dependent (dashed blue lines)
  and the result of the
   overall fit (solid blue line) is $w\approx 0.5$, see eq.\ (\ref{eq:wfit}).

\subsection*{Quadrupolar elastic interaction}

If two colloidal disks are embedded, boundary
conditions on the disk surface induce deformations in the
director field and, thus, induce an elastic interaction
given by the energy  (\ref{eq_frank_oseen_energy2D})
of the deformation. 
There are some  results on these elastic interactions in 
the one-constant approximation (${K = K_1 = K_3}$).
For two disks with perfect
planar boundary conditions there is an approximate
analytical result 
which is the   quadrupolar long-range interaction
from eq.\ (\ref{eq:quadrupolar_energy})
\cite{tasinkevych2002, dolganov2006}.
Since this is a field theoretical result, the needles are assumed to be
infinitely small relative to the colloids.
There are no analytical results for the full two-constant
free energy (\ref{eq_frank_oseen_energy2D}).

Therefore, we performed numerical simulations
based on the finite element method (using MATHEMATICA), i.e.,
solving eq.\ (\ref{eq:PDE}) 
for a system with two disks with distance $r$ and with an  angle $\vartheta$
with respect to the director axis  and using
the same anisotropy $K_1/K_3=0.1$ and weak anchoring strength
$w=0.5$ that we determined numerically.
We place both disks symmetrically in a system in a
square box with $L/\sigma=10$ and with a
director axis  $\Phi =0$  oriented 
parallel to one  edge of the simulation box
as fixed by Dirichlet conditions
at the system boundaries. 
The total energy of the system is the sum of elastic and
anchoring energy (see eq.\ (\ref{eq:F})).
The interaction energy  $U_\text{quad}(r, \vartheta)$
is obtained as the difference in energy
between a system containing two disks and the non-interacting system
as obtained by the sum of the energies of two systems containing
only one of the disks each. 
This results in Fig.\ \ref{fig_quadrupolar_interaction}(b).

\subsection*{Density-dependent depletion interaction}

We use the results of Biben {\it et al.}\cite{biben1996} and generalize them
to anisotropic depletants with a rotational degree of freedom $\varphi$ to get
a density-dependent depletion interaction for disks in a suspension of
hard needles.
We consider a system of hard disks with positions $\{ \vec{X}_I \}$
and $N_\text{n}$ hard needles with positions $\{ \vec{x}_i \}$ and
orientations $\{ \varphi_i \}$. Upper case indices refer to  disks,
lower case indices to needles.
The  energy of the system is given by
\begin{align*}
  H =
     \sum_{I<J} V_\text{dd}(\vec{X}_I - \vec{X}_J)
  +   \sum_{i<j} V_\text{nn}(\vec{x}_i - \vec{x}_j, \varphi_i, \varphi_j)
  +   \sum_{iI} V_\text{dn}(\vec{x}_i - \vec{X}_I, \varphi_i)
  \, .
\end{align*}
The disk-disk interaction is given by
$V_\text{dd}$, the needle-needle interaction by $V_\text{nn}$ and the
disk-needle interaction by $V_\text{dn}$.  By integrating over the needle
degrees of freedom one can derive the effective part of the
 interaction
${\mathcal V} (\{ \vec{X}_I \})$  between the
disks~\cite{biben1996},
\begin{align*}
  \beta \mathcal V ( \{ \vec{X}_I \} )
  = - \ln 
  \left[  \int \prod_i \mathrm d \vec{x}_i \mathrm d \varphi_i
    \exp  \left(  - \beta
      \left[  \sum_{iI} V_{dn}(\vec{x}_i - \vec{X}_I, \varphi_i)
        +
        \sum_{i<j} V_{nn}(\vec{x}_i - \vec{x}_j, \varphi_i \varphi_j)
      \right]
    \right)
  \right]
\end{align*}
($\beta\equiv 1/ k_\text{B} T$).
The corresponding
force $\mathcal F_K ( \{ \vec{X}_I \} )$ on disk $K$ is given by
\begin{align*}
  \mathcal F_K ( \{ \vec{X}_I \} )
  &=
  - \nabla_{\vec{X}_K}  \mathcal V ( \{ \vec{X}_I \} )
 =  -  \frac{1}{N_\text{n}}  \sum_l \int
  \rho^{(1)}(\vec{x}_l, \varphi_l | \{ \vec{X}_I \})
  \nabla_{\vec{X}_K}
  V_\text{dn}(\vec{x}_l - \vec{X}_K, \varphi_l)
  \mathrm d \vec{x}_l \mathrm d \varphi_l.
\end{align*}  
Here, we 
 used the single particle density of
needles with angle $\varphi_l$ at $\vec{x}_l$ for
fixed disk positions:
\begin{align*}
  \rho^{(1)}(\vec{x}_l, \varphi_l | \{ \vec{X}_I \})
  =
  N_\text{n}
  \frac{
    \int \prod_{i \neq l} \mathrm d \vec{x}_i \mathrm d \varphi_i
    \exp  \left(  - \beta
      \left[
        \sum_{iI} V_\text{dn}(\vec{x}_i - \vec{X}_I, \varphi_i)
        +
        \sum_{i<j} V_\text{nn}(\vec{x}_i - \vec{x}_j, \varphi_i, \varphi_j)
      \right]    \right)
  }{
    \int \prod_i \mathrm d \vec{x}_i \mathrm d \varphi_i
    \exp  
    \left(
      - \beta
      \left[
        \sum_{iI} V_\text{dn}(\vec{x}_i - \vec{X}_I, \varphi_i)
        +
        \sum_{i<j} V_\text{nn}(\vec{x}_i - \vec{x}_j, \varphi_i, \varphi_j)
      \right]
    \right)
  }
  \, .
\end{align*}
By using
${\nabla_{\vec{X}_K} V_\text{dn}(\vec{x_l}-\vec{X_K}, \varphi_l) = -
  \nabla_{\vec{x}_l} V_\text{dn}(\vec{x_l}-\vec{X_K}, \varphi_l)}$, evaluating
the sum to a factor $N_\text{n}$ and defining the average over the needle
angles as ${\langle A \rangle_\varphi = \int \mathrm d \varphi A(\varphi)}$,
we obtain 
\begin{align*}
  \mathcal F_{\vec{r}} ( \vec{0}, \vec{r} )
  &=
 \left\langle
  \int
  \rho^{(1)}(\pvec{r}, \varphi | \vec{0}, \vec{r})
  \nabla_{\pvec{r}}
  V_\text{dn}(\pvec{r} - \vec{r}, \varphi)
  \mathrm d \pvec{r}
  \right\rangle_\varphi
\end{align*}
for the case of two disks at $\vec{0}$ and $\vec{r}$.
We use the superposition approximation 
${\rho^{(1)}(\pvec{r}, \varphi | \vec{0}, \vec{r}) \approx \rho(\pvec{r},
  \varphi | \vec{0})\rho(\pvec{r}-\vec{r}, \varphi | \vec{0}) /
  \rho_\text{n}}$,
 where ${\rho(\pvec{r}, \varphi) \equiv \rho(\pvec{r},
  \varphi | \vec{0})}$ is the density distribution around a single
disk.
 For a single disk
the needles are distributed according to the direct interaction potential
$V_\text{dn}(\vec{r}, \varphi)$, 
\begin{align*}
  \rho(\vec{r}, \varphi)
  &=
    \rho_\text{n} \exp\left( - \beta V_\text{dn}(\vec{r}, \varphi) \right).
\end{align*}
This  finally leads to
\begin{align*}
  \mathcal F_{\vec{r}} ( \vec{0}, \vec{r} )
  &\approx
  \nabla_{\vec{r}}
    \frac{1}{\rho_\text{n} \beta}
    \left\langle
    \int
    \rho(\pvec{r}, \varphi)
    \rho(\pvec{r}-\vec{r}, \varphi)
    \mathrm d \pvec{r}
    \right\rangle_\varphi
    =
 - \nabla_{\vec{r}}U_\text{dep}(\vec{r}) \, .
\end{align*}
This effective potential is the  density-dependent depletion interaction,
which we further approximate by
employing a factorization approximation for the angular averages, 
${\langle\rho(\pvec{r}, \varphi) \rho(\pvec{r}-\vec{r}, \varphi)
  \rangle_\varphi \approx \langle \rho(\pvec{r}, \varphi) \rangle_\varphi
  \langle \rho(\pvec{r}-\vec{r}, \varphi) \rangle_\varphi}$, which is valid
for the isotropic phase and the ideal nematic phase.  Since we investigate the
effective interaction in the nematic phase this
should be a good approximation.
This gives our final result
\begin{align*}
  \beta U_\text{dep}(\vec{r})
  & \approx
  -
  \frac{1}{\rho_\text{n}}
  \int
  \langle  \rho(\pvec{r}, \varphi) \rangle_\varphi
  \langle  \rho(\pvec{r}-\vec{r}, \varphi)  \rangle_\varphi
  \mathrm d \pvec{r}
  =
  -  \rho_\text{n}
  \int
  \left(    1 - \frac{\rho(\pvec{r})}{\rho_\text{n}}  \right)
  \left(    1 - \frac{\rho(\pvec{r}-\vec{r})}{\rho_\text{n}}  \right)
  \mathrm d \pvec{r}.
\end{align*}
More intermediate steps of the derivation are given in the Supplementary
Information. 

\newpage


\section*{Acknowledgements}
We  acknowledge financial support by the Deutsche Forschungsgemeinschaft (DFG)
(grant No. KA 4897/1-1). {We also
  acknowledge financial support by Deutsche Forschungsgemeinschaft
  and TU Dortmund University within the funding programme
  Open Access Publishing.}

\section*{Author contributions statement}
D.M., T.K. and J.K. conceived and designed theory and  simulations. 
D.M. and T.K. performed the simulations.
 D.M., T.K. and J.K. analyzed the data and wrote the paper.

\section*{Supplemental Material}

\subsection*{Effective interaction and  residual   elastic interaction}

\begin{figure}[!h]
    \centering
    \includegraphics{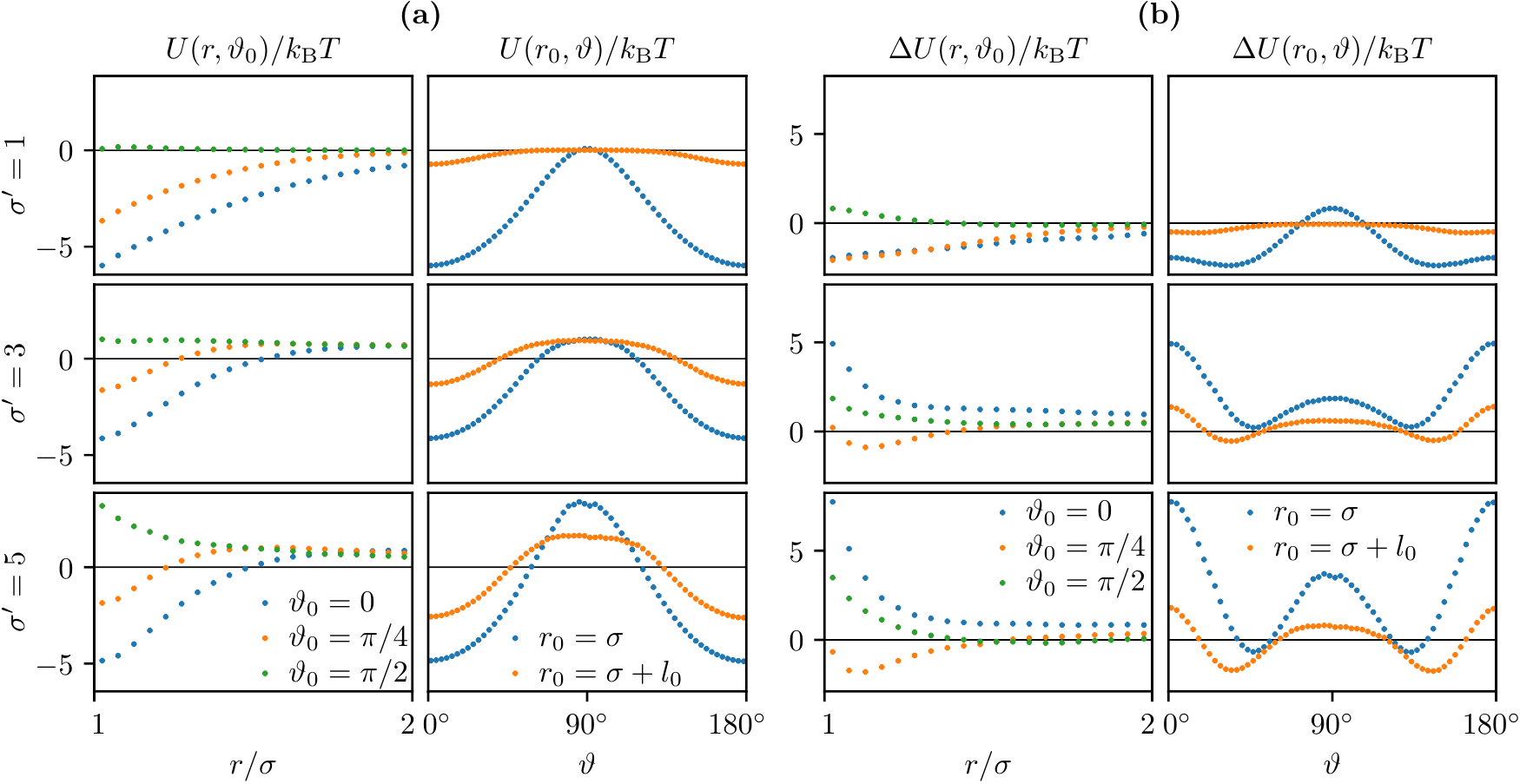}
    \caption{
      \textbf{(a)} Measured effective interaction of two hard disks
      suspended in hard needles for different disk diameters $\sigma'$.  The
      interaction is mostly attractive with a strength 
      around ${\sim 5 k_\text{B}T}$.  The interaction minimum is at the disks'
      surface at ${\vartheta = 0\degree}$.  The range of 
      ${U(r, \vartheta_0)}$ in radial direction
        (left column) is of order of $l_0$.
        In angular direction (right column),
        ${U(r_0, \vartheta)}$  shows a
      growing repulsive area at ${\vartheta=90\degree}$ for larger disks.
      \textbf{(b)} Residual elastic  interaction
      ${\Delta U (\vec{r}) = U(\vec{r}) - U_\text{dep}(\vec{r})}$ of two disks
      for different diameters $\sigma'$.  For larger disks the
       interaction shows a
      distorted quadrupolar pattern.  For small disks
      (${\sigma'=1}$) the repulsive area around ${\vartheta=0\degree/90\degree}$
      is missing. 
    }
    \label{fig_potentials_supp}
  \end{figure}

Fig.\ \ref{fig_potentials_supp} supplements Fig.\ 3 from the main text
and gives additional information on the  effective interaction $U$ and the
residual elastic interaction 
${\Delta U = U - U_\text{dep}}$ between two disks by showing
cuts through the contour plot in  Fig.\ 3 as a function of $r/\sigma$
or $\vartheta$. 

  
\subsection*{Event-chain algorithm}

The event-chain
  algorithm is a  rejection-free MC technique, which is based on
  global balance by introducing so-called lifting moves. 
  For hard spheres or needles a lifting move is the transfer of
  a MC displacement from one particle to another particle. 
  This means in a MC move only one particle at a
  time is active and moved along a line until it contacts another object.
  Then the remaining MC move distance is lifted to this object,
  which is then moved further.
  
  Needles are represented by their two endpoints, where only one endpoint 
  is moving at a time. In two dimensions the needle-needle interaction 
  simplifies effectively to a collision of an endpoint with another needle.
  The remaining MC move distance is
  lifted to one of the endpoints of this needle.
  Therefore, we have a fluid of endpoints with an
    effective 3-particle interaction
  (two endpoints of a passive needle and 
  the active endpoint). In Ref.\ \cite{harland2017}
   the generalization of the event-chain algorithm to N-particle interactions
   has been worked out. For needles the probability to which
   endpoint the MC move is lifted is proportional to the
   the distance to the other endpoint, i.e., it is lifted with
   higher probability to the closer endpoint.
   In the presence of additional disks, MC displacement
 is also lifted  to disks if a needle collides with the disk
 and vice versa.

\begin{figure}
    \centering
      \includegraphics{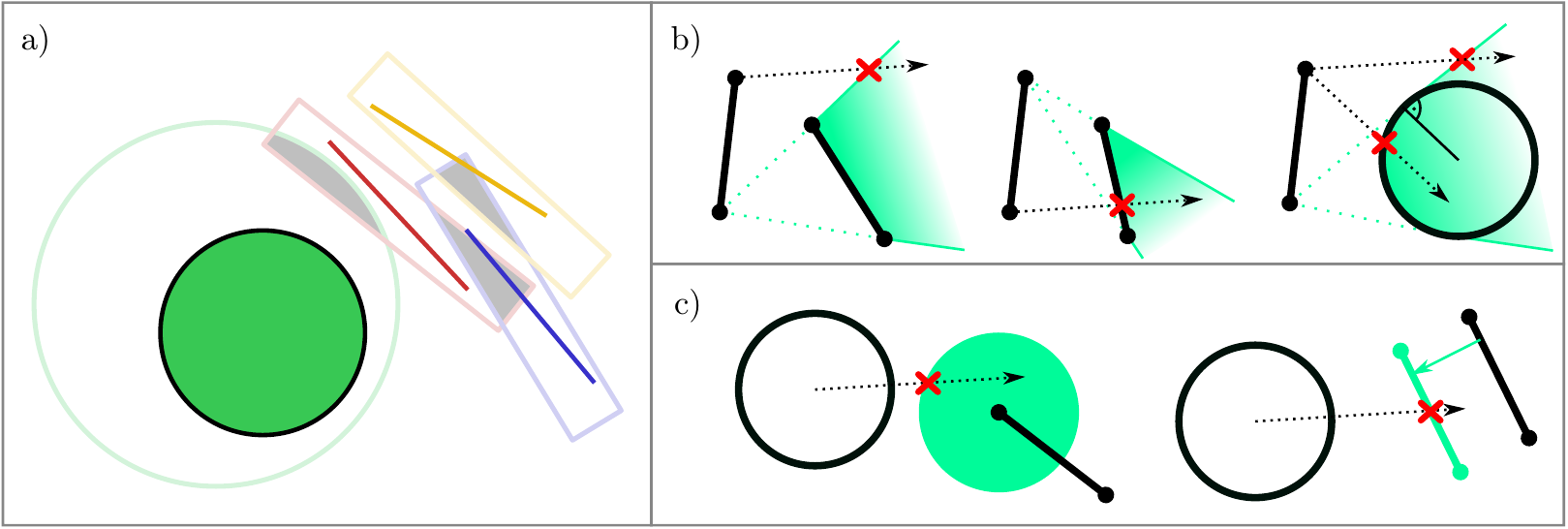}
      \caption{ \textbf{(a)} Scheme for the container neighbor lists.
        The figure shows three
        needles and a sphere and their respective containers.  Particle(s) are
        added to the neighbor lists of the other particle(s) and vice versa
        if their containers overlap. Arbitrary shapes address the anisotropy
        of needles efficiently.
        \textbf{(b)} Possible events in case a needle tip moves
        along the dashed black ray. Green lines are constructed by the
        inactive endpoint of the active needle and the endpoints of the hit
        needle or by a tangent to a hit sphere. Solid lines can trigger an
        event and the dashed part is to illustrate the construction.
        \textbf{(c)}
        Possible events in case a sphere moves along the dashed black ray.
        Left picture shows the handling of hitting an endpoint directly
        resembling the sphere-sphere interaction.  To check for the collision
        with the interior of the needle, one translates the needle
        perpendicular to its orientation by $\sigma /2$.  }
    \label{fig_neighbor_list}
  \end{figure}

   Effective collision detection is essential for a fast event-chain
    simulation. 
  The collisions are calculated by intersections of a ray starting at the
  active particle (either a sphere center or a needle endpoint) with another
  ray, line segment, or circle. The construction and an overview of all
  possible cases is shown  in Fig.\ \ref{fig_neighbor_list} (a) and (b).
  To speed
  up the collision detection, we use a special neighbor list design. Each
  particle is confined to a ``container'', which triggers an event when the
  particle leaves it.  Then the neighbor list is updated, which ensures that
  the neighbor lists are always valid. Particles are added to the neighbor
  lists of the other particle and vice versa when their containers
  overlap. This way, for different particles different container shapes can be
  chosen. For the needles a very narrow rectangle can be used, which limits
  the computational effort for calculating the distances to the next collision
  and makes the simulation significantly more efficient in the nematic
  phase. In particular, the anisotropy of the needles can be assigned
  particularly well without sacrificing any flexibility for the bookkeeping of
  the spheres. Even systems with a density of $\rho_\text{n}' = 100$,
  i.e., with a mean distance of $1/100 \l_0$, (see
  Fig.\ 2 in the main text) are possible.

  To avoid numerical issues we exclude the particle (needle or sphere) which
  was lifted from when calculating the rejection distance.  Furthermore, we
  optimize the updating of lists by putting them onto a collision grid.
  
  \subsection*{Derivation of the density-dependent depletion interaction}
  
  Here we present
  the derivation of the density-dependent depletion interaction
  including more 
  intermediate steps.
  We use the results of Biben {\it et al.}\cite{biben1996} and generalize them
  to anisotropic depletants with a rotational degree of freedom $\varphi$ to get
  a density-dependent depletion interaction for disks in a suspension of
  hard needles.
  We consider a system of hard disks with positions $\{ \vec{X}_I \}$
  and $N_\text{n}$ hard needles with positions $\{ \vec{x}_i \}$ and
  orientations $\{ \varphi_i \}$.  Upper case indices refer to  disks, s
  lower case indices to needles.
  The  energy of the system is given by
  \begin{align*}
    H =
      \sum_{I<J} V_\text{dd}(\vec{X}_I - \vec{X}_J)
    +   \sum_{i<j} V_\text{nn}(\vec{x}_i - \vec{x}_j, \varphi_i, \varphi_j)
    +   \sum_{iI} V_\text{dn}(\vec{x}_i - \vec{X}_I, \varphi_i)
    \, .
  \end{align*}
  The disk-disk interaction is given by
  $V_\text{dd}$, the needle-needle interaction by $V_\text{nn}$ and the
  disk-needle interaction by $V_\text{dn}$.  By integrating over the needle
  degrees of freedom one can derive the effective interaction
  ${\mathcal V (\{ \vec{X}_I \})}$  between the
  disks~\cite{biben1996},
  \begin{align*}
    \beta \mathcal V ( \{ \vec{X}_I \} )
    = - \ln 
    \left[  \int \prod_i \mathrm d \vec{x}_i \mathrm d \varphi_i
      \exp  \left(  - \beta
        \left[  \sum_{iI} V_{dn}(\vec{x}_i - \vec{X}_I, \varphi_i)
          +
          \sum_{i<j} V_{nn}(\vec{x}_i - \vec{x}_j, \varphi_i \varphi_j)
        \right]
      \right)
    \right]
  \end{align*}
  ($\beta\equiv 1/ k_\text{B} T$).
  The corresponding
  force $\mathcal F_K ( \{ \vec{X}_I \} )$ on disk $K$ is given by
  \begin{align*}
    \mathcal F_K ( \{ \vec{X}_I \} )
    &=
    - \nabla_{\vec{X}_K}  \mathcal V ( \{ \vec{X}_I \} )
    \\
    &=-   \sum_{l} 
    \int\left[\int \prod_{i \neq l} \mathrm d \vec{x}_i \mathrm d \varphi_i
      \exp  \left(
        - \beta
        \left[
          \sum_{iI} V_\text{dn}(\vec{x}_i - \vec{X}_I, \varphi_i)
          +
          \sum_{i<j} V_\text{nn}(\vec{x}_i - \vec{x}_j, \varphi_i, \varphi_j)
        \right]
      \right)
    \right]
    \\
    & \times
    \left[
      \int \prod_i \mathrm d \vec{x}_i \mathrm d \varphi_i
      \exp  
      \left(
        - \beta
        \left[
          \sum_{iI} V_\text{dn}(\vec{x}_i - \vec{X}_I, \varphi_i)
          +
          \sum_{i<j} V_\text{nn}(\vec{x}_i - \vec{x}_j, \varphi_i, \varphi_j)
        \right]
      \right)
    \right]^{-1}
    \\
    & \times
    \nabla_{\vec{X}_K} V_\text{dn}(\vec{x}_l - \vec{X}_K, \varphi_l)
    \mathrm d \vec{x}_l \mathrm d \varphi_l
    \\
    &=
    -
    \frac{1}{N_\text{n}}
    \sum_l \int
    \rho^{(1)}(\vec{x}_l, \varphi_l | \{ \vec{X}_I \})
    \nabla_{\vec{X}_K}
    V_\text{dn}(\vec{x}_l - \vec{X}_K, \varphi_l)
    \mathrm d \vec{x}_l \mathrm d \varphi_l.
  \end{align*}
  In the last step we used the single particle density of
  needles with angle $\varphi_l$ at $\vec{x}_l$ for
  fixed disk positions:
  \begin{align*}
    \rho^{(1)}(\vec{x}_l, \varphi_l | \{ \vec{X}_I \})
    =
    N_\text{n}
    \frac{
      \int \prod_{i \neq l} \mathrm d \vec{x}_i \mathrm d \varphi_i
      \exp  
      \left(
        - \beta
        \left[
          \sum_{iI} V_\text{dn}(\vec{x}_i - \vec{X}_I, \varphi_i)
          +
          \sum_{i<j} V_\text{nn}(\vec{x}_i - \vec{x}_j, \varphi_i, \varphi_j)
        \right]
      \right)
    }{
      \int \prod_i \mathrm d \vec{x}_i \mathrm d \varphi_i
      \exp  
      \left(
        - \beta
        \left[
          \sum_{iI} V_\text{dn}(\vec{x}_i - \vec{X}_I, \varphi_i)
          +
          \sum_{i<j} V_\text{nn}(\vec{x}_i - \vec{x}_j, \varphi_i, \varphi_j)
        \right]
      \right)
    }
    \, .
  \end{align*}
  By using
  ${\nabla_{\vec{X}_K} V_\text{dn}(\vec{x_l}-\vec{X_K}, \varphi_l) = -
    \nabla_{\vec{x}_l} V_\text{dn}(\vec{x_l}-\vec{X_K}, \varphi_l)}$, evaluating
  the sum to a factor $N_\text{n}$ and defining the average over the needle
  angles as ${\langle A \rangle_\varphi = \int \mathrm d \varphi A(\varphi)}$,
  we get
  \begin{align*}
    \mathcal F_K ( \{ \vec{X}_I \} )
    &=
    \frac{1}{N_\text{n}}
    \sum_l \int
    \rho^{(1)}(\vec{x}_l, \varphi_l | \{ \vec{X}_I \})
    \nabla_{\vec{x}_l}
    V_\text{dn}(\vec{x}_l - \vec{X}_K, \varphi_l)
    \mathrm d \vec{x}_l \mathrm d \varphi_l
    \\
    &=
    \left\langle
    \int
    \rho^{(1)}(\pvec{r}, \varphi | \{ \vec{X}_I \})
    \nabla_{\pvec{r}}
    V_\text{dn}(\pvec{r} - \vec{X}_K, \varphi)
    \mathrm d \pvec{r}
    \right\rangle_\varphi
    \, .
  \end{align*}
  For the case of two disks at $\vec{0}$ and $\vec{r}$ this yields
  \begin{align}
    \mathcal F_{\vec{r}} ( \vec{0}, \vec{r} )
    &=
    \left\langle
    \int
    \rho^{(1)}(\pvec{r}, \varphi | \vec{0}, \vec{r})
    \nabla_{\pvec{r}}
    V_\text{dn}(\pvec{r} - \vec{r}, \varphi)
    \mathrm d \pvec{r}
    \right\rangle_\varphi
    \label{eq:F0r}
  \end{align}
  We use the superposition approximation
  \begin{align*}
  {\rho^{(1)}(\pvec{r}, \varphi | \vec{0}, \vec{r}) \approx \rho(\pvec{r},
    \varphi | \vec{0})\rho(\pvec{r}-\vec{r}, \varphi | \vec{0}) /
    \rho_\text{n}},
  \end{align*}
  where ${\rho(\pvec{r}, \varphi) \equiv \rho(\pvec{r},
  \varphi | \vec{0})}$ is the density distribution around a single
  disk and $\rho_\text{n}$ is the average needle density. For a single disk
  the needles are distributed according to the direct interaction potential
  $V_\text{dn}(\vec{r}, \varphi)$, 
  \begin{align*}
    \rho(\vec{r}, \varphi)
    &=
    \rho_\text{n} \exp\left( - \beta V_\text{dn}(\vec{r}, \varphi) \right) 
  \end{align*}
  resulting in 
  \begin{align*}
    \nabla_{\vec{r}}
    \rho(\vec{r}, \varphi)
    &=
    - \beta
      \rho_\text{n} \exp\left( - \beta V_\text{dn}(\vec{r}, \varphi) \right)
    \nabla_{\vec{r}} V_\text{dn}(\vec{r}, \varphi)
    =
    - \beta
    \rho(\vec{r}, \varphi)
    \nabla_{\vec{r}} V_\text{dn}(\vec{r}, \varphi) \, .
  \end{align*}
  Using this in eq.\ (\ref{eq:F0r})
  we arrive at
  \begin{align*}
    \mathcal F_{\vec{r}} ( \vec{0}, \vec{r} )
    &\approx
    - \frac{1}{\rho_\text{n} \beta}
    \left\langle
    \int
    \rho(\pvec{r}, \varphi)
        \rho(\pvec{r}-\vec{r}, \varphi)
        (- \beta \nabla_{\pvec{r}}
        V_\text{dn}(\pvec{r} - \vec{r}, \varphi))
    \mathrm d \pvec{r}
    \right\rangle_\varphi
    \\
    &=
    \nabla_{\vec{r}}
      \frac{1}{\rho_\text{n} \beta}
      \left\langle
      \int
      \rho(\pvec{r}, \varphi)
      \rho(\pvec{r}-\vec{r}, \varphi)
      \mathrm d \pvec{r}
      \right\rangle_\varphi
      =
      - \nabla_{\vec{r}}U_\text{dep}(\vec{r}) \, .
  \end{align*}
  This effective potential is the  density-dependent depletion interaction,
  which we further approximate by
  \begin{align*}
    \beta U_\text{dep}(\vec{r})
    & \approx
    -
    \frac{1}{\rho_\text{n}}
    \int
    \langle  \rho(\pvec{r}, \varphi) \rangle_\varphi
    \langle    \rho(\pvec{r}-\vec{r}, \varphi) \rangle_\varphi
    \mathrm d \pvec{r}
    =
    -    \frac{1}{\rho_\text{n}}
    \int \rho(\pvec{r}) \rho(\pvec{r}-\vec{r})
    \mathrm d \pvec{r}
    \\
    &=
    -
    \rho_\text{n}
    \int
    \left(
      1 - \frac{\rho(\pvec{r})}{\rho_\text{n}}
    \right)
    \left(
      1 - \frac{\rho(\pvec{r}-\vec{r})}{\rho_\text{n}}
    \right)
    \mathrm d \pvec{r} \, ,
  \end{align*}
  where we used
  ${\langle\rho(\pvec{r}, \varphi) \rho(\pvec{r}-\vec{r}, \varphi)
    \rangle_\varphi \approx \langle \rho(\pvec{r}, \varphi) \rangle_\varphi
    \langle \rho(\pvec{r}-\vec{r}, \varphi) \rangle_\varphi}$, which is valid
  for the isotropic phase and the ideal nematic phase.
  Since we investigate the
  effective interaction in the nematic phase this
  should be a good approximation.
  
\begin{figure}
    \centering
    \includegraphics{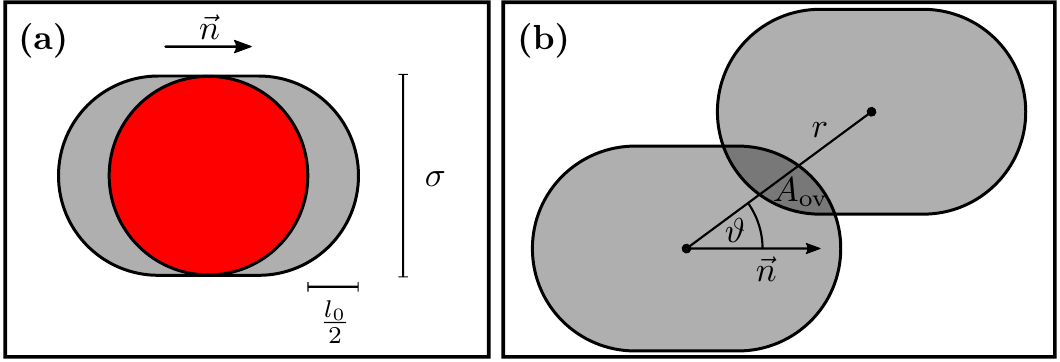}
    \caption{
      Idealized step-like depletion zone around a disk (a) and
      resulting overlap area $A_\text{ov}(r, \vartheta)$ (b).
    }
    \label{fig:Aov}
  \end{figure}

  For the special case of an idealized
  density that is a step function and either zero or $\rho_\text{n}$
  (see Fig.\ \ref{fig:Aov}), the
  effective potential essentially becomes the well-known depletion interaction
  $\beta U(r)=-\rho_0 A_\text{ov}$, where $A_\text{ov}$ is the
  overlap area of the excluded areas~\cite{asakura1954}.

\end{document}